\documentclass[prd,nofootinbib,showpacs,showkeys]{revtex4}

\usepackage{mathrsfs}
\usepackage{tipa}
\usepackage{bm}
\usepackage{hyperref}

\usepackage{amsmath}
\usepackage{amssymb}
\usepackage{amsthm}
\usepackage{amsfonts}
\usepackage{mathtools}
\usepackage{latexsym}
\usepackage{relsize}
\usepackage{booktabs}
\usepackage{tikz}
\usetikzlibrary{positioning}
\usepackage[letterpaper,margin=0.5in]{geometry}
\usepackage{lscape}

\usepackage{array}
\newcolumntype{?}{!{\vrule width 2pt}}

\newcommand{\lp}{\left(}
\newcommand{\rp}{\right)}
\newcommand{\lb}{\left[}
\newcommand{\rb}{\right]}

\newcommand{\lsim}   {\mathrel{\mathop{\kern 0pt \rlap
  {\raise.2ex\hbox{$<$}}}
  \lower.9ex\hbox{\kern-.190em $\sim$}}}
\newcommand{\gsim}   {\mathrel{\mathop{\kern 0pt \rlap
  {\raise.2ex\hbox{$>$}}}
  \lower.9ex\hbox{\kern-.190em $\sim$}}}
\newcommand{\bw}{\begin{widetext}\begin{equation}}
\newcommand{\ew}{\end{equation}\end{widetext}}
\newcommand{\be}{\begin{equation}}
\newcommand{\ee}{\end{equation}}
\newcommand{\ba}{\begin{eqnarray}}
\newcommand{\ea}{\end{eqnarray}}

\newcommand{\diff}{{{\rm d}}}

\newcommand{\X}{\mathcal{X}}

\newcommand{\nn}{\nonumber}

\begin{document}

\title{Noether symmetries in Symmetric Teleparallel Cosmology}
\author{Konstantinos F. Dialektopoulos}
\email{dialektopoulos@na.infn.it}
\affiliation{Center for Gravitation and Cosmology, College of Physical Science and Technology, Yangzhou University, Yangzhou 225009, China,}
\affiliation{Aristotle University of Thessaloniki, 54124 Thessaloniki, Greece,}
\affiliation{Institute of Space Sciences and Astronomy, University of Malta, Msida, MSD 2080, Malta}

\author{Tomi S. Koivisto}
\email{timoko@kth.se}
\affiliation{Nordita, KTH Royal Institute of Technology and Stockholm University, Roslagstullsbacken 23, 10691 Stockholm, Sweden,}
\affiliation{Laboratory of Theoretical Physics, Institute of Physics, University of Tartu, W. Ostwaldi 1, 50411 Tartu, Estonia,} 
\affiliation{ National Institute of Chemical Physics and Biophysics, R\"avala pst. 10, 10143 Tallinn, Estonia}

\author{Salvatore Capozziello}
\email{capozzie@na.infn.it}
\affiliation{Dipartimento di Fisica, Universit\'a di Napoli {}``Federico II'', Compl. Univ. di Monte S. Angelo, Edificio G, Via Cinthia, I-80126, Napoli, Italy,}
\affiliation{INFN Sezione  di Napoli, Compl. Univ. di Monte S. Angelo, Edificio G, Via Cinthia, I-80126, Napoli, Italy.}
\affiliation{Gran Sasso Science Institute, Via F. Crispi 7, I-67100, L'Aquila, Italy,}
\affiliation{Tomsk State Pedagogical University, ul. Kievskaya, 60, 634061 Tomsk, Russia.}

\preprint{NORDITA-2019-048}
\date{\today}

\begin{abstract}
We consider a general theory of all possible quadratic, first-order derivative terms of the non-metricity tensor in the framework of Symmetric Teleparallel Geometry. We apply the Noether Symmetry Approach to classify those models that are invariant under point transformations in a cosmological background and we use the symmetries of these models to reduce the dynamics of the system in order to find analytical solutions. 

\end{abstract}

\maketitle


\section{Introduction}

General Relativity (GR) and the associated cosmological model, i.e. $\Lambda$ Cold Dark Matter ($\Lambda$CDM) model have passed several  observational tests, but are plagued with some shortcomings as well. The difficulty to find out a convincing explanation for the today observed small value of the cosmological constant, to detect  particle candidates for Dark Matter, to incorporate early inflation and late accelerated expansion  in a self-consistent  paradigm, as well as to  formulate  a quantum  theory of gravity, in order to  embed $\Lambda$CDM into the Standard Model of particles, are only some of them. One more recent fallacy appeared, after scientists measured locally the value of the Hubble constant $H_0$ \cite{Riess:2019cxk}, and compared it to its value inferred from Planck CMB observations \cite{planck1,planck2}. The discrepancy presented between these two measurements is of $4.4\sigma$ in significance, making the evidence for physics beyond $\Lambda$CDM more plausible. In order to tackle all these issues, either partially or as a whole, scientists started to study modifications and extensions of GR \cite{Capozziello:2002rd,Clifton:2011jh,Capozziello:2011et,Nojiri:2017ncd}.

Furthermore, it is well-known that GR has two more equivalent descriptions, that are based on different connections. Specifically, GR is formulated using the Levi-Civita connection, that is induced by the spacetime metric and thus gravity is mediated through curvature, in a torsion-free spacetime with vanishing non-metricity. If we consider a metric compatible but flat connection (i.e. the curvature is zero), we obtain the Teleparallel Equivalent of General Relativity (TEGR), where gravity is mediated through torsion. Finally, one can consider a flat and torsion-free connection in order to formulate the Symmetric Teleparallel Equivalent of GR (STEGR) where non-metricity is the property that mediates gravitational interactions.  These three equivalent descriptions are often euphemistically called ``The Geometrical Trinity of Gravity'' \cite{BeltranJimenez:2019tjy}.

At the level of field equations, all these three theories are completely equivalent, meaning that all experimental tests satisfied by GR, will be satisfied by the other two theories as well. However, since they are formulated in different geometries, once we modify them, their modifications need not necessarily be equivalent. At a fundamental level, the three different  approaches could constitute an important arena to test the Equivalence Principle in its various formulations \cite{Altschul:2014lua}. 

In this paper, we will take into account the  Symmetric Teleparallel Geometry, where non-metricity is non-zero, while torsion and curvature vanish. Recently, new classes of extended gravity theories have been explored in such a geometrical setting \cite{BeltranJimenez:2017tkd,Conroy:2017yln,Jarv:2018bgs,Runkla:2018xrv,Harko:2018gxr,Adak:2018vzk,Hohmann:2018wxu,Soudi:2018dhv,Iosifidis:2018zwo}. We will consider all the possible quadratic terms of the non-metricity tensor, $Q_{\alpha\mu\nu} = \nabla_{\alpha}g_{\mu\nu}\,,$ that are first order in the derivatives of the metric; there are five of these and we name them $A,B,C,D$ and $E$ (there is one more possible term, which however is parity-violating and trivial in the cosmological background). The most general theory with these scalars is given by an arbitrary function of them, i.e. $f(A,B,C,D,E)$. We shall consider the cosmology of this very generic class of theories, and after constructing its point-like Lagrangian, we will apply the so-called Noether Symmetry Approach \cite{Capozziello:1994du,Capozziello:1996bi, Capozziello:1996ay,Capozziello:2009te,Dialektopoulos:2018qoe} to classify those models that are invariant under point transformations. 

Noether point symmetries are a subclass of Lie symmetries applied to dynamical systems that are described by a point Lagrangian and leave the action integral invariant. There is a well known method \cite{Capozziello:2018gms,Bahamonde:2016grb,Bahamonde:2016jqq,Capozziello:2016eaz,Basilakos:2013rua,Bahamonde:2017sdo} that can be used as a geometric criterion to constrain theories of gravity. Moreover, by using the symmetries of each theory, we can reduce the dynamics of its system in order to find out analytical solutions.

The paper is organized as follows: in Sec.\ref{sec:geometry} we present the geometric setup of the geometrical trinity of gravity and in particular of the STEGR. In addition, we discuss in greater detail modifications of STEGR and specifically $f(Q)-$models. In Sec.\ref{sec:f(A,B,C,D,E)} we present the general 2nd order models, dubbed $f(A,B,C,D,E)$ and we discuss its cosmology. In Sec.\ref{sec:comf_sym} we present some models that are invariant under infinitesimal diffeomorphisms (Diffs) and also covariant under the conformal transformation $g_{\mu\nu}\rightarrow e^{2\phi}g_{\mu\nu},\,\Gamma ^{\alpha}{}_{\mu\nu} \rightarrow \Gamma ^{\alpha}{}_{\mu\nu}$. In Sec.\ref{sec:Noether_classification}, after introducing the Noether Symmetry Approach, we apply it to the point-like cosmological Lagrangian of $f(A,B,C,D,E)$ and we present a classification of the invariant models. Finally, in Sec.\ref{sec:Noether_conformal},  we study the Noether symmetries of a specific model that is conformally invariant and we use its symmetries to find exact cosmological solutions. In Sec.\ref{sec:Discussion} we discuss our results and present future perspectives.  

\section{The geometric setup}
\label{sec:geometry}

It is known from differential geometry \cite{Hehl:1994ue} that a generic affine connection can be decomposed into the following three parts
\begin{equation}\label{gen_affine_connection}
\Gamma ^{\alpha}{}_{\mu\nu} = \left\{^{\phantom{i} \alpha}_{\mu\nu}\right\} + K^{\alpha}{}_{\mu\nu} + L ^{\alpha}{}_{\mu\nu}\,.
\end{equation}
The first term is the known Levi-Civita connection that reads
\be \label{christoffel}
\left\{^{\phantom{i} \alpha}_{\beta\gamma}\right\} \equiv \frac{1}{2}g^{\alpha\lambda}\lp g_{\beta\lambda,\gamma}
+  g_{\lambda\gamma,\beta} - g_{\beta\gamma,\lambda}\rp\,,
\ee
the second and third terms are the  contorsion and disformation\footnote{The effect of non-metricity upon the affine connection is called disformation because, in addition to rescaling, the non-metric transformations include shear transformations as well. The combined effect of contorsion and disformation is sometimes called distortion in the literature, and contorsion is often called contortion instead.} tensors respectively, that are defined as
\begin{align}\label{contorsion}
K^{\alpha}{}_{\mu\nu} &= \frac{1}{2}g^{\alpha\beta}\left(T_{\mu\beta\nu}+T_{\nu\beta\mu}+T_{\beta\mu\nu}\right)\,,\\ \label{disformation}
L^{\alpha}{}_{\mu\nu} &= \frac{1}{2}g^{\alpha\beta}\left(Q_{\beta\mu\nu}-Q_{\mu\beta\nu}-Q_{\nu\beta\mu}\right)\,.
\end{align}
The torsion and the non-metricity tensors are given by
\begin{gather}\label{torsion_tensor}
T^{\alpha}{}_{\mu\nu} \equiv \Gamma ^{\alpha}{}_{\mu\nu} - \Gamma ^{\alpha}{}_{\nu\mu}\,, \\ \label{nonmetricity_tensor}
Q_{\alpha\mu\nu} \equiv \nabla_{\alpha}g_{\mu\nu} = \partial _{\alpha}g_{\mu\nu} - \Gamma ^{\kappa}{}_{\alpha\mu}g_{\kappa\nu} - \Gamma ^{\alpha}{}_{\alpha\nu}g_{\mu\kappa}\,.
\end{gather}
There is one more property of the connection, its curvature, and it is defined as
\begin{equation}\label{riemann}
R^{\alpha}{}_{\beta\mu\nu} \equiv 2 \partial _{[\mu}\Gamma ^{\alpha}{}_{\nu ]\beta} + 2 \Gamma ^{\alpha}{}_{[ \mu | \lambda | }\Gamma ^{\lambda}{}_{\nu ]\beta}\,.
\end{equation}

In  teleparallel geometry, the connection is constrained by ${R}^\alpha_{\phantom{\alpha}\beta\mu\nu} = 0$. Then
$\Gamma^\alpha_{\phantom{\alpha}\mu\nu}$, has the form of an inertial general linear transformation, which further reduces to the Weitzenb\"ock form, once the structure group is restricted to the (pseudo)rotational one. In symmetric teleparallel geometry, the latter restriction is abandoned. Instead, the torsion is constrained to vanish, $T^\alpha_{\phantom{\alpha}\mu\nu}=0$. This leaves the inertial connection in the form of a pure translation.

Let us then relate the conventional formulation of GR in the geometry \eqref{christoffel} to the symmetric teleparallel geometry.
We denote the derivative with respect to the Christoffel symbol \eqref{christoffel} by $\mathcal{D}_\alpha$, so that $\mathcal{D}_\alpha g_{\mu\nu}=0$. For a general affine connection however, i.e. \eqref{gen_affine_connection}, we can define the non-metricity tensor as \eqref{nonmetricity_tensor}. This tensor has two independent traces, which we denote as $Q_\mu=Q_{\mu\phantom{\alpha}\alpha}^{\phantom{\mu}\alpha}$ and $\tilde{Q}_\mu = Q_{\alpha\mu}{}^\alpha$. The Weyl connection represents the prototype non-metricity 
\be
W^{\alpha}_{\phantom{\alpha}\mu\nu}  \equiv  \frac{1}{2}g_{\mu\nu}Q^\alpha-\delta^\alpha_{(\mu}Q_{\nu)}\,. \label{disformation2b}
\ee
We note that since $L^\alpha_{\phantom{\alpha}\mu\alpha} = -\frac{1}{2}Q_\mu$ and $W^\alpha_{\phantom{\alpha}\mu\alpha} = -\frac{n}{2}Q_\mu$,
the combination losing one trace in $n$ dimensions is $L^\alpha_{\phantom{\alpha}\mu\nu} - \frac{2}{n}W^\alpha_{\phantom{\alpha}\mu\nu}$. 
However, we will be most interested in the non-metricity scalar
\be \label{qdef}
Q \equiv \frac{1}{2}Q_{\alpha\beta\gamma} \lp L^{\alpha\beta\gamma} - W^{\alpha\beta\gamma}\rp\,.
\ee
To see the relevance of this quadratic form, we first note that when shifting a connection as
$\hat{\Gamma}^\alpha_{\phantom{\alpha}\mu\nu} = {\Gamma}^\alpha_{\phantom{\alpha}\mu\nu} + \Omega^\alpha_{\phantom{\alpha}\mu\nu}$, 
the curvature of the shifted connection can be written as  
\be 
 \hat{R}^\alpha_{\phantom{\alpha}\beta\mu\nu}  =  R^\alpha_{\phantom{\alpha}\beta\mu\nu} + T^\lambda_{\phantom{\lambda}\mu\nu}\Omega^\alpha_{\phantom{\alpha}\lambda\beta}  +  2\nabla_{[\mu}\Omega^\alpha_{\phantom{\alpha}\nu]\beta} +2\Omega^\alpha_{\phantom{\alpha}[\mu\lvert\lambda\rvert}\Omega^\lambda_{\phantom{\lambda}\nu]\beta}\,. \label{hatR}
\ee
With $R$ the scalar curvature of (\ref{riemann}), and $\mathcal{R}$ the scalar curvature of metrical connection (\ref{christoffel}), we obtain, from the above formula,
the following identity (where the first equality holds when there is no torsion and the second when there is no curvature):
\be \label{ricciscalarq}
R   =  \mathcal{R}  + Q +   \mathcal{D}_\alpha ( Q^\alpha - \tilde{Q}^\alpha ) = 0\,.
\ee
Thus, the dynamics of the Einstein-Hilbert theory $2\mathcal{L} = \mathcal{R}$ in the Riemannian geometry is reproduced by the Lagrangian $2\mathcal{L}=-Q$ in the symmetric
teleparallel geometry. In fact the latter is an improved version of the theory since it has no second derivatives hiding inside a boundary term.

\subsection{The $f(Q)$ models}

Let us  consider the class of models that generalizes the GR-equivalent $2\mathcal{L}=-Q$ by a function $2\mathcal{L}=-f(Q)$. 
Two variational methods have been considered recently: 
\begin{itemize}
\item The Palatini variation \cite{BeltranJimenez:2017tkd,BeltranJimenez:2018vdo}. The variational degrees of freedom are the metric $g_{\mu\nu}$ and the affine connection $\Gamma^\alpha{}_{\mu\nu}$. The action is then $S_G  =   \int \diff^n x\lb-\frac{1}{2}\sqrt{-g} f(Q)
 +  \lambda_\alpha^{\phantom{\alpha}\beta\mu\nu} R^\alpha_{\phantom{\alpha}\beta\mu\nu} + \lambda_\alpha^{\phantom{\alpha}\mu\nu}T^\alpha_{\phantom{\alpha}\mu\nu}\rb\,.$
The desired geometry is set with Lagrange multiplier (tensor densities) and a technical complication in the complete analysis is that one needs to obtain the solutions also for these fields.
\item The inertial variation \cite{Golovnev:2017dox}. The variational degrees of freedom are the metric $g_{\mu\nu}$ and the translation that generates the inertial connection. The
action can be written just as $S_G  =   -\frac{1}{2}\int \diff^n x\sqrt{-g} f(Q)$, since now the variations are by construction restricted to the torsion-free and flat geometry. 
\end{itemize}
It could also be useful to consider a combination of the above methods\footnote{We thank Jos\'e{} Beltr\'a{}n for pointing this out.}:
 \begin{itemize}
\item Assume the connection to be flat as in Ref.\cite{Golovnev:2017dox}, and then impose its symmetry by a Lagrange multiplier $S_G  =   \int \diff^n x\lb-\frac{1}{2}\sqrt{-g} f(Q) + \lambda_\alpha^{\phantom{\alpha}\mu\nu}T^\alpha_{\phantom{\alpha}\mu\nu}\rb\,.$ The variational degrees of freedom are the metric and the general linear transformation that generates a flat connection. A benefit of this formulation with respect to the previous one is that no higher derivatives of the St\"uckerberg field appear in the action. This method facilitate the study of the generic (allowing both torsion and non-metricity) teleparallel gravity theories, which however is not the purpose of the present paper. 
\end{itemize}
The metric field equations for this theory are 
\be
{T}_{\mu\nu} =  f''Q_{,\alpha}\lp L^\alpha_{\phantom{\alpha}\mu\nu} - W^\alpha_{\phantom{\alpha}\mu\nu}\rp   +  f'\lb \lp\mathcal{D}_\alpha-\frac{1}{2}Q_\alpha\rp L^\alpha_{\phantom{\alpha}\mu\nu} + \frac{1}{2}\mathcal{D}_\mu Q_{\nu} - L^\alpha_{\phantom{\alpha}\beta\mu}L^\beta_{\phantom{\beta}\alpha\nu}\rb - \frac{1}{2}g_{\mu\nu}\lb f + f'\mathcal{D}_\alpha\lp Q^\alpha-\tilde{Q}^\alpha\rp \rb\,,\label{fqees}
\ee
where ${T}_{\mu\nu}$ is the matter energy momentum tensor.
In general we need the field equations also for the connection, and for that the second and third method above is probably the more convenient one. 

Since we can understand gravitation as a gauge theory, we can consider gauges which may be less conventional than the Levi-Civita one but, at least for some purposes, are more convenient. It is interesting to consider the possible simplification of the
theory in the so-called ``coincident'' gauge, $\Gamma^\alpha_{\phantom{\alpha}\mu\nu}=0$ \cite{BeltranJimenez:2017tkd}.

In cosmology we can neglect the inertial connection at the background level, since there exist consistent solutions with the gauge choice $\Gamma^\alpha{}_{\mu\nu}=0$. In a cosmology described by the line element $\diff s^2 = -\diff t^2 + a^2(t)\delta_{ij}\diff x^i \diff x^j$, the expansion rate can be defined as $H=\dot{a}/a$. The field equations (\ref{fqees}) 
adapted to this line element are
\begin{align}
6f'H^2-\frac{1}{2}f & =  8\pi G\rho\,, \label{fried1} \\
\lp 12f''H^2 + f'\rp\dot{H} & =  -4\pi G\lp\rho+p\rp\,. \label{fried2}
\end{align}
We have  restored the Newton constant $G$ and included a matter source with the energy density $\rho$ and the pressure $p$. The continuity can be verified immediately, $\dot{\rho}=-3H(\rho+p)$, which is a cross-check of the consistency of our gauge choice. It turns out that Friedmann equations equivalent to above in some power-law $f(Q)$-models were already studied  in the framework of ``vector distortion'' cosmology \cite{Jimenez:2016opp}. More precisely, the expansion histories of the $f(Q)\sim Q \pm M^4/Q$ and the  $f \sim Q \pm (Q/m)^2$ cases (with $m$ and $M$ some mass scales) coincide with those in the cosmologies of some particular  quadratic gravity models with ``vector distortion''. In fact, the equations \eqref{fried1},\eqref{fried2} are the same as in the
teleparallel $f(T)$ cosmology \cite{Cai:2015emx}, and the boundary term is also equivalent at the level of the cosmological background \cite{Bahamonde:2016grb}. So there are plenty of known solutions to these equations \cite{Cai:2015emx}.

\section{General second order models: $f(A,B,C,D,E,F)$}
\label{sec:f(A,B,C,D,E)}

Thus, it is known about symmetric teleparallel cosmology that the Friedmann equations of $f(Q)$ can reproduce those of $f(T)$. 
As a much more generic action, one could consider some
function of all the six quadratic invariants 
\be\label{letters1}
A \equiv  Q_{\alpha\mu\nu}Q^{\alpha\mu\nu}\,, \quad B  \equiv Q_{\alpha\mu\nu}Q^{\mu\alpha\nu}\,, \quad C \equiv  Q_\alpha Q^\alpha , \quad D  \equiv  \tilde{Q}_\alpha \tilde{Q}^\alpha\,, \quad E  \equiv  \tilde{Q}_\alpha {Q}^\alpha\,, \quad F \equiv \epsilon^{\alpha\beta\mu\nu}Q_{\alpha\beta}{}^\rho Q_{\mu\nu\rho}\,,
\ee
minimally coupled to matter with a Lagrangian $\mathcal{L}_\text{m}$,  
\be \label{action}
S = \int \diff^4 x \sqrt{-g}\lb f(A,B,C,D,E,F) + \mathcal{L}_\text{m}\rb\,. 
\ee 
We note that the function $f$ defines the constitutive relation for the non-metricity tensor,
\be \label{nmsuper}
P^\alpha{}_{\mu\nu}  \equiv \frac{\partial f}{\partial Q_\alpha{}^{\mu\nu}}\,.
\ee
The metric field equations, coupled to the source $T_{\mu\nu} = \delta \lp \sqrt{-g}\mathcal{L}_{\text{m}}\rp/\delta g^{\mu\nu}$, are 
\be \label{geom}
\frac{2}{\sqrt{-g}}\nabla_\alpha \lp\sqrt{-g}P^\alpha{}_{\mu\nu}\rp - \frac{\partial f}{\partial g^{\mu\nu}} -  \frac{1}{2}f g_{\mu\nu}= T_{\mu\nu}\,.
\ee
Even if the affine connection vanishes everywhere, it can have physical effects and thus its equation of motion is
\be \label{ceom}
\nabla_\mu\nabla_\nu \lp \sqrt{-g}P^{\mu\nu}{}_\alpha\rp = 0\,.
\ee

\subsection{Cosmology}

Let us begin using the second method. We then need to include a lapse function $N(t)$ into the cosmological line element,
\be \label{nfrw}
\diff s^2 = -N^2(t)\diff t^2 + a^2(t)\delta_{ij}\diff x^i \diff x^j\,.
\ee
In addition to the expansion rate $H=\dot{a}/a$ we can define the dilation rate $T=\dot{N}/N$.  
The non-vanishing connection coefficients (\ref{christoffel}) are 
\be
\left\{^{\phantom{i} 0}_{00}\right\}  =   T\,, \quad  \left\{^{\phantom{i} 0}_{ij}\right\}  =   \frac{a^2H}{N^2}\delta_{ij}\,,  \quad \left\{^{\phantom{i} i}_{0j}\right\}  =   \left\{^{\phantom{i} i}_{j0}\right\} = H\delta^i_j\,,
\ee
and the only non-vanishing components of the non-metricity tensor are 
\be
Q_{000}  =   -2N^2T \,, \quad Q_{0ij} =   2a^2H\delta_{ij}\,,
\ee
while the traces are
\be
Q_\alpha  =  2\lp 3H+T\rp\delta_\alpha^0\,, \quad \tilde{Q}_\alpha =  2T\delta_\alpha^0\,.
\ee
We thus easily obtain the quadratic invariants for the metric \eqref{nfrw}:
\begin{subequations}
\label{letters}
\begin{align}
A & \equiv  Q_{\alpha\mu\nu}Q^{\alpha\mu\nu} = -4N^{-2}\lp 3H^2 + T^2\rp\,, \\
B & \equiv  Q_{\alpha\mu\nu}Q^{\mu\alpha\nu} = -4N^{-2}T^2\,, \\
C & \equiv  Q_\alpha Q^\alpha = -4N^{-2}\lp 3H+T\rp^2\,, \\
D & \equiv  \tilde{Q}_\alpha \tilde{Q}^\alpha = -4N^{-2}T^2\,, \\
E & \equiv  \tilde{Q}_\alpha {Q}^\alpha = -4N^{-2}\lp 3H + T\rp T\,, \\
F & \equiv \epsilon^{\alpha\beta\mu\nu}Q_{\alpha\beta}{}^\rho Q_{\mu\nu\rho} =0\,.
\end{align}
\end{subequations}
Since only the even-parity invariants are non-trivial in the cosmological background, we will discard the $F$ from further consideration.
We can then check that since, from \eqref{qdef},
\be
Q = \frac{1}{4}A - \frac{1}{2}B - \frac{1}{4}C + \frac{1}{2}E = 6N^{-2} H^2\,,
\ee
and for the boundary term we have
\be
 \mathcal{D}_\alpha ( Q^\alpha - \tilde{Q}^\alpha ) = -6N^{-2}\lp \dot{H}+3H^2-TH\rp\,,
\ee
we obtain the correct expression for the Ricci scalar from the relation \eqref{ricciscalarq},
\be
\mathcal{R}   =   -Q - \mathcal{D}_\alpha ( Q^\alpha - \tilde{Q}^\alpha ) =    6N^{-2}\lp \dot{H} + 2H^2 -TH\rp\,. 
\ee
The action \eqref{action} can then be put into the form
\be \label{action2}
S = \int a^3 N\lb f(A,B,C,D,E) + \mathcal{L}_m\rb \diff t \equiv  \int L \diff t  \,.
\ee
In the following, we shall consider the action in terms of the metric variables $N$ and $a$.
To this end, we rewrite the action \eqref{action2} as
\begin{align}
S  =   &\int a^3 N\lb f(A,B,C,D,E) + \mathcal{L}_m\rb -   \lambda_A\lb A + 4N^{-2}\lp 3H^2+T^2\rp\rb - \nn \\
 &-    \lambda_B\lp B + 4N^{-2}T^2\rp  -   \lambda_C\lb C + 4N^{-2}\lp 3H+T\rp^2\rb  -   \lambda_D\lp D + 4N^{-2}T^2\rp -   \lambda_E\lb E + 4N^{-2}\lp 3H+T\rp T\rb\,. 
\end{align}
Let us denote the five invariants collectively with $X$, i.e.  $X \in \{A,B,C,D,E\}$. Varying with respect to each $X$ we obtain the five equations
\be
a^3 N f_{X} - \lambda_X = 0\,. 
\ee
The Lagrange multipliers $\lambda_X$ can therefore be straightforwardly solved, and we can eliminate them from the action to obtain
\be
S  =   \int \diff t\Big\{a^3 N\lb f(A,B,C,D,E) + \mathcal{L}_m\rb  
 -   \frac{a^3}{N}\sum_X f_{X} \lp N^2 X + 4T^2\rp 
 -  \frac{12a^3}{N}\lb  \lp f_{A} + 3f_{C}\rp H^2 + \lp 2f_{C}+f_{E}\rp H T \rb\Big\}\,. \label{point}
\ee 
Matter we consider is a perfect fluid in the cosmological background (\ref{nfrw}), so that\footnote{With this convention, our action $S$ is in units of $S_{\text{can}}/16\pi G$, where
$S_{\text{can}}$ is the action with the canonical dimension and $G$ is the Newton constant. Since constant rescalings of the action are irrelevant to the dynamics, we can ease the notation with this convention.}
\be
\mathcal{L}_m = 16\pi G \rho\,, \quad \rho \sim a^{-3(1+w)}\,.
\ee
Here $w$ is interpreted as the equation of state, and $\rho$ as the energy density of matter. For simplicity we take $w$ to be a constant, which is sufficient to cover the two most relevant cases, a universe filled with the dust, $w=0$, and a universe filled with radiation, $w=1/3$.

The field equations then follow from the Euler-Lagrange equations as
\begin{align} \label{e-la}
\frac{\diff}{\diff t} \frac{\partial L}{\partial \dot{a}}  &=  \frac{\partial L}{\partial a}\,,  \\ \label{e-ln}
\frac{\diff}{\diff t} \frac{\partial L}{\partial \dot{N}}  &=  \frac{\partial L}{\partial N}\,, 
\end{align}
and they come with the energy condition
\be
\label{e-c}
\frac{\partial L}{\partial \dot{a}}\dot{a} + \frac{\partial L}{\partial \dot{N}}\dot{N} + \sum_{X}\frac{\partial L}{\partial \dot{X}}\dot{X}=L\,.  
\ee
The equation of motion for the lapse function \eqref{e-ln} yields the generalised Friedmann equation,
\begin{align}
8\pi G N^2\rho =   &\frac{N^2}{2} f + 6\lp 2f_C+ f_E\rp\dot{H}  +   6\lp 2f_A + 12f_C+3f_E\rp H^2 + 4\sum_X f_X \dot{T}  +   6\lp 2f_A+4f_C+2f_B+2f_D+3f_E\rp HT +\nn \\
& +  \sum_X\lb 6\lp 2f_{CX}  +  f_{EX}\rp H +4\sum_Y f_{XY} T\rb \dot{X}\,. \label{f1a}
\end{align}
The other Friedmann equation, for the scale factor \eqref{e-la}
is
\begin{align}
- 8\pi G N^2 w\rho  =   &\frac{N^2 }{2} f + 4\lp f_A+3f_C\rp \lp \dot{H} + 3H^2-HT\rp +   2\lp 2f_C + f_E\rp \lp \dot{T}-T^2 + 3HT\rp  \nn \\
& +  4\sum_X\Big[ 2\lp f_{AX}+3f_{CX}\rp H  
 +   \lp 2f_{CX} + f_{EX}\rp T\Big] \dot{X} \,. \label{f2}
\end{align}
The energy condition \eqref{e-c} imposes a nontrivial constraint that relates the two expansion rates by
\be
-2N^2\rho  = N^2\lp f-\sum_X f_X X\rp 
 +  12\lp f_A+3f_C\rp H^2 + 12\lp 2f_C+f_E\rp HT + 4\sum_X f_X T^2\,. \label{f1b}
\ee
This completes the field equations for the most general symmetric teleparallel one-derivative theory in flat cosmological background.
In fact it would be quite feasible to allow higher derivatives, i.e. $\dot{X}$-dependence of the point-like Lagrangian, but 
in this exploratory study we are content with the function of the five $X$'s.  As a cross-check, in the case of $f(Q)$, we have
$4f_A = -2f_B =-4f_C = 2f_E = f'(Q)$ and $F_0 \sim F_1 \sim F_2 \sim F_3 \sim f''(Q)$, and can verify that both  \eqref{f1a} and \eqref{f1b} reduce to \eqref{fried1}, whilst 
\eqref{f2} reduces to \eqref{fried2}.

\section{Conformal Symmetries}
\label{sec:comf_sym}

It is of our interest to focus on some more specific classis of theories that enjoy enhanced gauge invariance. That is the reason why in this section we consider those models that are invariant under conformal and infinitesimal diffeomorphic transformations. 

\subsection{Infinitesimal diffeomorphisms}

The quadratic theory that retains its invariance in the coincident gauge under infinitesimal Diffs is given by
\begin{equation}
Q_{\text{Diff}} = c_1 A+c_2 B- c_1 C - (2c_1+c_2)D+2 c_1 E\,. \nonumber
\end{equation}
Setting $c_2 = -2 c_1$ we are left with the nonlinear invariant theory $Q_{\text{Diff}} = -4c_1 Q$. We have already seen that the Friedman-Robertson-Walker cosmology does not distinguish between the models $f(Q_{\text{Diff}})$ and $f(Q)$. One may however consider the wider class of theories which is invariant only under transverse Diffs. These are given by the 4-parameter case
\cite{BeltranJimenez:2018vdo}
\begin{equation}
Q_{\text{TDiff}} = c_1 A+c_2 B +c_3 C - (2 c_1 + c_2) D + c_5 E \,.
\end{equation}
we may further require the invariance under Weyl rescalings. This fixes one of the parameters in the above as \cite{BeltranJimenez:2018vdo}
\begin{equation}
Q_{\text{WTDiff}} = c_1 A + c_2 B - \frac{3}{8} c_1 C - (2 c_1 + c_2) D + c_1 E\,.
\end{equation}

\subsection{Metric rescalings}

There are three quadratic combinations that transform covariantly under the conformal rescaling $g_{\mu\nu} \rightarrow e^{2\phi}g_{\mu\nu}\,,$ $\Gamma ^{\alpha}{}_{\mu\nu} \rightarrow \Gamma ^{\alpha}{}_{\mu\nu}\,.$ These are \cite{Iosifidis:2018zwo}
\begin{equation} \label{invars}
\hat{A} \equiv A+4D-2E\,,\quad \hat{B}\equiv B-D\,,\quad \hat{C}\equiv C+16 D-8E\,. \nonumber
\end{equation}
The general scale-invariant theory is quartic in non-metricity and therefore has six parameters:
\begin{equation}
f = a_1 \hat{A}^2 + a_2 \hat{B}^2 + a_3 \hat{C}^2 + a_4 \hat{A}\hat{B} + a_5 \hat{A}\hat{C}+ a_6 \hat{B}\hat{C}\,.
\end{equation}
On the other hand, we can consider the class of quadratic theories which describes a general scale-invariant scalar-nonmetricity theory in a fixed gauge (the scalar is set to a constant that fixes the Newton's constant). Then we have
\begin{equation}\label{fnew}
f = b_1 \hat{A} + b_2 \hat{B} +b_3 \hat{C} + b_4 C + b_5\,.
\end{equation}
The parameter $b_4$ comes from the kinetic term of the scalar and the cosmological constant $b_5$ from the quartic potential that is allowed by the scale invariance.

It may be useful to clarify the motivation of this theory in more detail, since in the following we shall focus on its Noether symmetries in the cosmological background. From Ref. \cite{Iosifidis:2018zwo} we see\footnote{In particular, see their Eq.(60). In the published version of  \cite{Iosifidis:2018zwo}
there is however a typo in Table 3, since their $\hat{A}$ and $\hat{C}$ differ from those in (\ref{invars}).} that the three Diff-invariants in (\ref{invars}) transform scale-covariantly $\hat{X} \rightarrow e^{-2\phi}\hat{X}$ under a conformal transformation of the metric $g_{\mu\nu} \rightarrow e^{2\phi}g_{\mu\nu}$ which leaves the connection unchanged (in particular, we can remain in the coincident gauge despite the rescaling). Since under this transformation $\sqrt{-g} \rightarrow e^{4\phi}\sqrt{-g}$, we may achieve an invariant action by considering terms such as $\sqrt{-g}\psi^2\hat{X}$, where
$\psi$ is a scalar field that has the proper conformal weight such that $\psi \rightarrow e^{-\phi}\psi$.  We may also introduce dynamics for this compensating scalar field, and for that purpose we
need to consider a scale-covariant derivative. Let us we follow Weyl's original notation and denote with an asterisk the scale covariant derivative
\be
\overset{\ast}{\nabla}_\mu\psi
\equiv (\partial_\mu - \frac{1}{8}Q_\mu)\psi\,.
\ee
Then we can write down the 5-parameter scale-invariant symmetric teleparallel action
\be
\hat{S} = \int \diff^4 x \sqrt{-g}\lb \psi^2\lp b_1 \hat{A} + b_2 \hat{B} +b_3 \hat{C} \rp + 64b_4 (\overset{\ast}{\nabla}_\mu\psi) (\overset{\ast}{\nabla}{}^\mu\psi) + b_5 \psi^4 +\mathcal{L}_m\rb\,.
\ee 
We have also taken into account the quartic self-interaction of the scalar field $\psi$ which is allowed by the conformal rescaling symmetry. We can now choose such a local rescaling 
$\phi=\log{\psi}$ of the fields that $\psi \rightarrow e^{-\phi}\psi = 1$, to obtain the action (\ref{action}) wherein the function $f$ is given by (\ref{fnew}). If matter is described by a constant $w$ as suggested above, one should restrict to $w=1/3$ in order not to break the invariance of the theory.

\section{Models Classification by the Noether Symmetry Approach}
\label{sec:Noether_classification}

Let us  briefly describe point transformations and the Noether Symmetry Approach; more details can be found in \cite{Capozziello:1996bi, Dialektopoulos:2018qoe}. Consider a set of one parameter point transformations of the form
\begin{equation}\label{point_transformations}
\tilde{t} = Z(t,q^i,\epsilon)\,,\quad \tilde{q}^i = \Gamma ^i(t,q^i,\epsilon)\,,
\end{equation}
where $t$ is the independent variable, $q^i$ the variable on the configuration space and $\epsilon$ the parameter of the transformation. We can construct the generator of the above transformation as
\begin{equation}\label{Noether_generator}
\X = \xi (t,q^i) \partial _t + \eta ^i (t,q^i) \partial _i\,,
\end{equation}
where 
\begin{equation}
\xi (t,q^i) = \frac{\partial Z(t,q^i,\epsilon)}{\partial \epsilon}|_{\epsilon \rightarrow 0} \quad \text{and} \quad \eta^i(t,q^i) = \frac{\partial \Gamma ^i(t,q^i,\epsilon)}{\partial \epsilon}|_{\epsilon \rightarrow 0}\,.
\end{equation}
In addition, the $n^{\text{th}}$ prolongation of the generator \eqref{Noether_generator} is
\begin{equation}
\X^{[n]} = \X + \eta ^i _{[1]} \partial _{\dot{q}^i} + ... + \eta ^i _{[n]} \partial _{q^{(n)i}}\,,
\end{equation}
with $\eta ^i _t = D_t \eta ^i - q^i D_t \xi\,,$ and $D_t = \frac{\partial}{\partial t} + \dot{q}^i \frac{\partial }{\partial \dot{q}^i}$.

Now let us consider a dynamical system described by the point-like Lagrangian $L = L(t,q^i,\dot{q}^i)$. The Euler-Lagrange (EL) equations are given by
\begin{equation}
E_i(L) = 0 \Rightarrow \frac{\diff}{\diff t}\frac{\partial L}{\partial \dot{q}^i} - \frac{\partial L}{\partial q^i} = 0\,.
\end{equation}
These equations are invariant under \eqref{Noether_generator} or \eqref{point_transformations} if and only if there exists a function $h = h(t,q^i)$ such that
\be
\X^{[1]}\mathcal{L} + \mathcal{L}\frac{d\xi}{dt} = \frac{dh}{dt}\,. \label{NoetherCondition}
\ee
In that case we say that $\X$ is a Noether symmetry of the dynamical system of $L$ and 
\begin{equation}
I = \xi(\dot{q}^i \frac{\partial L}{\partial \dot{q}^i}-L) - \eta ^i \frac{\partial L}{\partial q^i} + h\,,
\end{equation}
is the first integral of the equations of motion, i.e. the associated conserved quantity of the symmetry $\X$.

Once we have found the symmetry of a theory, i.e. the non-zero coefficients of the Noether vector $\X$, we can construct a Lagrange system of the form
\begin{equation}
\frac{dt}{\xi} = \frac{\diff q^i}{\eta ^i} = \frac{\diff \dot{q}^i}{\eta^{[1]i}} \,,
\end{equation}
in order to find the zero-th and first order invariants
\begin{equation}
W^{[0]}(t,q^i) = \frac{\diff t}{\xi} - \frac{\diff q^i}{\eta^i} \quad \text{and} \quad W^{[1]}(t,q^i,\dot{q}^i) = W^{[0]} - \frac{\diff \dot{q}^i}{\eta ^{[1]i}}\,.
\end{equation}
Using these invariants one can reduce the dynamics of the system, i.e. the order of the EL equations and solve the simplified system to find exact solutions.

In our case, we choose an \textit{ansatz} for the generator \eqref{Noether_generator} of the form
\be
\X = \xi \partial_t + \eta ^a\partial _a+ \eta ^N\partial _N+ \eta ^\chi\partial _{\chi}\,, \label{generator}
 \ee
where we denote collectively all the quadratic invariants \eqref{nfrw} as $\chi$ and the coefficients of the vector $(\xi,\eta^i)$ depend on $(t,a,N,\chi)$. $\X^{[1]}$ is the first prolongation of \eqref{generator} and in our case, it  is given by 
\begin{align}
\X^{[1]} &= \X + \left(\dot{\eta}^i - \dot{\xi}\dot{q}^i \right)\partial_{\dot{q}^i}= \X + \left(\dot{\eta}^a - \dot{\xi}\dot{a} \right)\partial_{\dot{a}}+\left(\dot{\eta}^N - \dot{\xi}\dot{N} \right)\partial_{\dot{N}}\,.
\end{align}
A note here is necessary: as we can immediately see, the point-like Lagrangian \eqref{point} does not depend on the derivative of the $\chi$ configuration variables. This means that in principle, one could solve the EL equations, i.e. $\partial L/\partial \chi = 0$ and substitute back to the Lagrangian to make it canonical. However, this complicates the calculations a lot and we avoid it. In addition, it has been pronven \cite{Havelkova,Zi-ping} that the Noether Symmetry Approach \eqref{NoetherCondition} is valid even for singular Lagrangians.

By applying the condition \eqref{NoetherCondition} to the Lagrangian \eqref{point} we get a system of 40 equations, which of course, are not all independent. What we do in this paper is  classifying different theories, i.e. different forms of the function $f(A,B,C,D,E)$, that present  Noether symmetries in the cosmological minisuperspace (see also \cite{Capozziello:2012hm} on this topic). 

Before proceeding with the classification, let us put things into a perspective: the condition \eqref{NoetherCondition} is an equation of the form $H(a,\dot{a},N,\dot{N},\chi,\dot{\chi},\xi,\dot{\xi},\eta^i,\dot{\eta}^i,f,\dot{f},\dot{g}) = 0$; or else, it is a polynomial of $\dot{q}^i$. In order for this polynomial to vanish, all the coefficients of $\dot{q}^i$ have to vanish identically. This is how we end up with a system of 40 equations. The unknowns are the 8 coefficients of the generating vector, $(\xi,\eta^i)$ and the functions $f(\chi)$ and $h(t,a,N,\chi)$. The system would be overdetermined if the equations would be all independent; however this is not the case. In order to be able to solve the system, we consider in all the following cases that, the function $h$ on the right hand side of the condition $\eqref{NoetherCondition}$ is constant. This is indeed an arbitrary constraint, however, it is the mildest assumption we can do to simplify the system, and eventually solve it. 

In this way, there are 12 different classes of theories that show symmetries. The classification can be seen in the following figure \ref{fig:M1} and the table \ref{tab:sum}. The different cases are denoted in the diagram, using the symbol ``C\textit{n.l}", where \textit{n} is a number and \textit{l} is a letter, while the classes of theories with Noether symmetries are denoted with \textbf{S\textit{n}}, where again, \textit{n} is a number. In the table \ref{tab:sym} we present the different classes of theories.
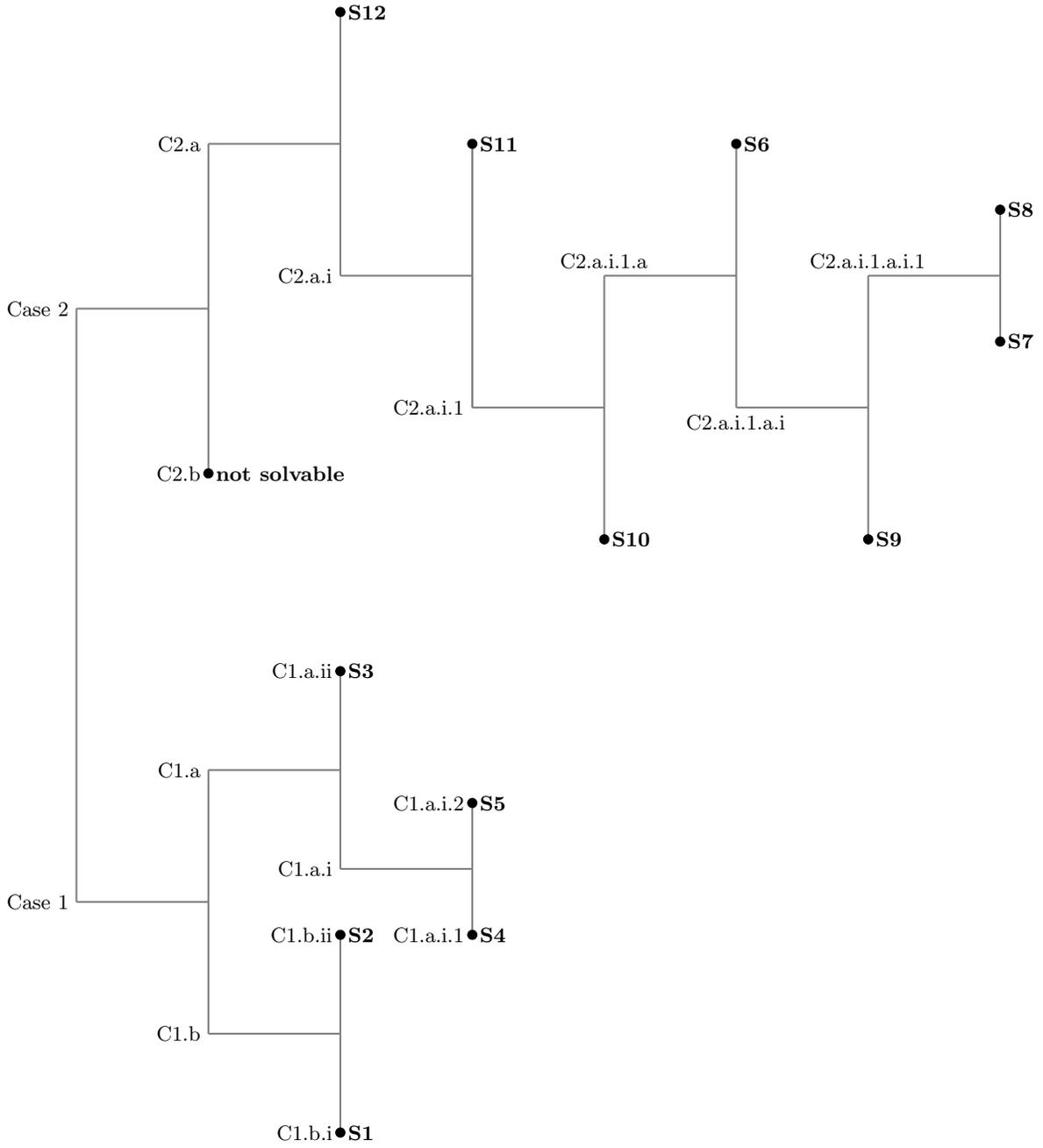
\begin{figure}[!h]
\begin{center}
\begin{tikzpicture}
\draw[gray, thick] (-3,-4.5) -- (-3,4.5);
\draw[gray, thick] (-3,-4.5) -- (-1,-4.5);
\draw[gray, thick] (-3,4.5) -- (-1,4.5);
\draw[gray, thick] (-1,-6.5) -- (-1,-2.5);
\draw[gray, thick] (-1,2) -- (-1,7);
\draw[gray, thick] (-1,-6.5) -- (1,-6.5);
\draw[gray, thick] (-1,-2.5) -- (1,-2.5);
\draw[gray, thick] (1,-4) -- (3,-4);
\draw[gray, thick] (1,5) -- (3,5);
\draw[gray, thick] (3,3) -- (5,3);
\draw[gray, thick] (5,5) -- (7,5);
\draw[gray, thick] (7,3) -- (9,3);
\draw[gray, thick] (9,5) -- (11,5);
\draw[gray, thick] (1,-8) -- (1,-5);
\draw[gray, thick] (1,-4) -- (1,-1);
\draw[gray, thick] (1,5) -- (1,9);
\draw[gray, thick] (3,-5) -- (3,-3);
\draw[gray, thick] (3,3) -- (3,7);
\draw[gray, thick] (3,3) -- (3,7);
\draw[gray, thick] (-1,7) -- (1,7);
\draw[gray, thick] (5,1) -- (5,5);
\draw[gray, thick] (7,3) -- (7,7);
\draw[gray, thick] (9,1) -- (9,5);
\draw[gray, thick] (11,4) -- (11,6);
\filldraw[black] (1,-8) circle (2pt) node[anchor=west] {\textbf{S1}};
\filldraw[black] (1,-5) circle (2pt) node[anchor=west] {\textbf{S2}};
\filldraw[black] (1,-1) circle (2pt) node[anchor=west] {\textbf{S3}};
\filldraw[black] (3,-5) circle (2pt)  node[anchor=west] {\textbf{S4}};
\filldraw[black] (3,-3) circle (2pt) node[anchor=west] {\textbf{S5}};
\filldraw[black] (-1,2) circle (2pt) node[anchor=west] {\textbf{not solvable}};
\filldraw[black] (1,9) circle (2pt) node[anchor=west] {\textbf{S12}};
\filldraw[black] (3,7) circle (2pt) node[anchor=west] {\textbf{S11}};
\filldraw[black] (7,7) circle (2pt) node[anchor=west] {\textbf{S6}};
\filldraw[black] (5,1) circle (2pt) node[anchor=west] {\textbf{S10}};
\filldraw[black] (9,1) circle (2pt) node[anchor=west] {\textbf{S9}};
\filldraw[black] (11,4) circle (2pt) node[anchor=west] {\textbf{S7}};
\filldraw[black] (11,6) circle (2pt) node[anchor=west] {\textbf{S8}};
\filldraw[black] (-3,-4.5)  node[anchor=east] {Case 1};
\filldraw[black] (-1,-6.5)  node[anchor=east] {C1.b};
\filldraw[black] (-1,-2.5)  node[anchor=east] {C1.a};
\filldraw[black] (1,-4)  node[anchor=east] {C1.a.i};
\filldraw[black] (3,-5)  node[anchor=east] {C1.a.i.1};
\filldraw[black] (3,-3)  node[anchor=east] {C1.a.i.2};
\filldraw[black] (1,-1)  node[anchor=east] {C1.a.ii};
\filldraw[black] (1,-8)  node[anchor=east] {C1.b.i};
\filldraw[black] (1,-5)  node[anchor=east] {C1.b.ii};
\filldraw[black] (-3,4.5)  node[anchor=east] {Case 2};
\filldraw[black] (-1,2)  node[anchor=east] {C2.b};
\filldraw[black] (-1,7)  node[anchor=east] {C2.a};
\filldraw[black] (1,5)  node[anchor=east] {C2.a.i};
\filldraw[black] (3,3)  node[anchor=east] {C2.a.i.1};
\filldraw[black] (5,5)  node[anchor=south] {C2.a.i.1.a};
\filldraw[black] (7,3)  node[anchor=north] {C2.a.i.1.a.i};
\filldraw[black] (9,5)  node[anchor=south] {C2.a.i.1.a.i.1};
\end{tikzpicture}
\end{center}
\caption{In this scheme we represent the diffent cases ``C\textit{n.l}" of each class of symmetries \textbf{S\textit{n}}. Each case and each symmetry are explicitly given in the table \ref{tab:sum}.} \label{fig:M1}
\end{figure}

\begin{table}
\renewcommand{\arraystretch}{1.5}
\centering
\centering
\resizebox{19cm}{!}{
\begin{tabular}{c|c}
\hline
\hline
\textbf{Cases} & \textbf{Conditions}   \\ 
\hline
\textbf{1} & $f(A,B,C,D,E)=f_1(A,B,D,-\frac{C}{2}+E)$ \\ 
\hline 
\textbf{2} &  $f(A,B,C,D,E) \neq f_1(A,B,D,-\frac{C}{2}+E)$\\
 \hline 
\textbf{C1.a} & $f_1(A,B,D,-\frac{C}{2}+E) = f_2(B,D,\frac{3A}{2}-\frac{C}{2}+E)$\\
 \hline 
 \textbf{C1.b} & $f_1(A,B,D,-\frac{C}{2}+E) \neq f_2(B,D,\frac{3A}{2}-\frac{C}{2}+E)$\\
 \hline 
 \textbf{C1.a.i} & $f_2(B,D,\frac{3A}{2}-\frac{C}{2}+E) \neq f_3(\frac{3A}{2}-2B-\frac{C}{2}+E,-B+D)$\\
 \hline 
 \textbf{C1.a.ii} & $f_2(B,D,\frac{3A}{2}-\frac{C}{2}+E) = f_3(\frac{3A}{2}-2B-\frac{C}{2}+E,-B+D)$\\
 \hline 
 \textbf{C1.b.i} & $f_1(A,B,D,-\frac{C}{2}+E) = f_2(-A+B,-A+D,-\frac{A}{2}-\frac{C}{2}+E)$\\
 \hline 
 \textbf{C1.b.ii} & $f_1(A,B,D,-\frac{C}{2}+E) \neq f_2(-A+B,-A+D,-\frac{A}{2}-\frac{C}{2}+E)$\\
 \hline 
 \textbf{C1.a.i.1} & $f_2(B,D,\frac{3A}{2}-\frac{C}{2}+E) = Bf_3(\frac{D}{B},\frac{3A-C+2E}{2B})$\\
 \hline 
 \textbf{C1.a.i.2} & $f_2(B,D,\frac{3A}{2}-\frac{C}{2}+E) \neq Bf_3(\frac{D}{B},\frac{3A-C+2E}{2B})$\\
 \hline
 \textbf{C2.a} & $f(A,B,C,D,E)=f_1(-A+B,-A+C,-A+D,-A+E)$\\
   \hline 
 \textbf{C2.b} & $f(A,B,C,D,E) \neq f_1(-A+B,-A+C,-A+D,-A+E)$\\
   \hline 
 \textbf{C2.a.i} & $f_1(-A+B,-A+C,-A+D,-A+E) = f_2(-3A+2B+C,-B+D,-B+E)+f_3(-B+C,-B+D,-B+E)$\\
   \hline 
 \textbf{C2.a.ii} & $f_1(-A+B,-A+C,-A+D,-A+E) \neq f_2(-3A+2B+C,-B+D,-B+E)+f_3(-B+C,-B+D,-B+E)$\\
   \hline 
 \textbf{C2.a.i.1} & $f_3(-B+C,-B+D,-B+E) = f_4(-B+D,-B+E) + f_5(-C+D,-C+E)$\\
   \hline
 \textbf{C2.a.i.2} & $f_3(-B+C,-B+D,-B+E) \neq f_4(-B+D,-B+E) + f_5(-C+D,-C+E)$\\
 \hline
 \textbf{C2.a.i.1.a} & $f_5(-C+D,-C+E) = f_6(-C+E) +f_7 (-D+E)$ \\
 \hline
 \textbf{C2.a.i.1.b} & $f_5(-C+D,-C+E) \neq f_6(-C+E) +f_7 (-D+E)$ \\
 \hline
 \textbf{C2.a.i.1.a.i} & $f_6(-C+E) \neq f_{6c}(-C+E) +f_{6cc}$ \\
 \hline
 \textbf{C2.a.i.1.a.ii} & $f_6(-C+E) = f_{6c}(-C+E) +f_{6cc}$ \\
  \hline
 \textbf{C2.a.i.1.a.i.1} & $f_2 = f_8(-B+D,-B+E) + f_9(-B+D,\frac{3A}{2}-2B-\frac{C}{2}+E)$ \\
  \hline
 \textbf{C2.a.i.1.a.i.2} & $f_2 \neq f_8(-B+D,-B+E) + f_9(-B+D,\frac{3A}{2}-2B-\frac{C}{2}+E)$ \\
  \hline
 \textbf{C2.a.i.1.a.i.1.i} & $f_4(-B+D,-B+E) = -f_8(-B+D,-B+E)$ \\
  \hline
 \textbf{C2.a.i.1.a.i.1.ii} & $f_4(-B+D,-B+E) \neq - f_8(-B+D,-B+E)$ \\
 \hline
 \hline
  \end{tabular}}
  \caption{In this table we show the different cases, i.e. ``C\textit{n.l}'' of the figure \ref{fig:M1}.}
  \label{tab:sum}
\end{table}

\begin{table}[!h]
\renewcommand{\arraystretch}{1.5}
\centering
\centering
\resizebox{20cm}{!}{
\begin{tabular}{c|c|c|c|c|c|c|c|c|c|c}
\hline
\hline
\textbf{Symmetry} & $\xi (t,a,N,X)$ & $\eta _a(t,a,N,X)$ & $\eta _N(t,a,N,X)$ & $\eta _A(t,a,N,X)$ & $\eta _B(t,a,N,X)$ & $\eta _C(t,a,N,X)$ & $\eta _D(t,a,N,X)$ & $\eta _E(t,a,N,X)$ & $h(t,a,N,X)$ & $f(X) $  \\ 
\hline
\textbf{S1} & $ \xi (t)$ & $\eta _a (a)$& known & known &arbitrary &arbitrary &arbitrary &arbitrary & $h_0$& $f(-A+B,-A+D,-\frac{A}{2}-\frac{C}{2}+E)$\\
 \hline
\textbf{S2} &$ \xi (t)$ & $\eta _a (a)$&$\eta_N(N)$ &known &known &known &arbitrary &arbitrary &$h_0$ & $f(A,B,D,-\frac{C}{2}+E)$\\
 \hline
\textbf{S3} & $ \xi (t)$ & known & arbitrary &arbitrary &arbitrary &arbitrary &arbitrary &arbitrary & $h(t)$& $f(\frac{3A}{2}-2B-\frac{C}{2}+E,-B+D)$ \\
 \hline
\textbf{S4} & $ \xi (t)$&known & $\eta_N(N)$& arbitrary & arbitrary  & arbitrary  & arbitrary  & arbitrary  &$h_0$ & $Bf(\frac{D}{B},\frac{3A-C+2E}{2B})$\\
 \hline
\textbf{S5} & $ \xi (t)$& known & $\eta_N(N)$&known & arbitrary  & arbitrary  & arbitrary  & arbitrary  & $h_0$& $f(B,D,\frac{3A}{2}-\frac{C}{2}+E)$\\
 \hline
\textbf{S6} &$\xi_1 t+ \xi_2$ & $\eta_{a1}a + \frac{1}{4}\eta_{N1}a^2$ &$\left( 3\eta_{a1}+\eta_{N1}a -\xi_1\right)N$ & arbitrary  &known & arbitrary  & arbitrary  & arbitrary  &$h_0$ & $f_{6c}\left(-\frac{21}{8}A+\frac{7}{4}B-\frac{1}{8}C+E\right)+f_{6cc}+f_8(-B+D,\frac{3A}{2}-2B-\frac{C}{2}+E)$ \\
 \hline
\textbf{S7} &$ \xi (t)$ &$\eta _a (a)$ &$\eta_N(a,N)$ &  arbitrary &known & known & arbitrary  &known &$h_0$ & $f_6(-C+E)+f_7(-D+E)+f_9(-B+D,\frac{3A}{2}-2B-\frac{C}{2}+E)$\\
 \hline
\textbf{S8} &$ \xi (t)$ &$\eta _a (a)$ & $\eta_N(a,N)$& arbitrary  &known &known & arbitrary  &known & $h_0$& $f_4(-B+D,-B+E)+f_6(-C+E)+f_7(-D+E)+f_8(-B+D,-B+E)+f_9(-B+D,\frac{3A}{2}-2B-\frac{C}{2}+E)$\\
 \hline
\textbf{S9} &$ \xi (t)$ & $\eta _a (a)$&$\eta_N(a,N)$ & known  & known  & arbitrary  & arbitrary  & known  & $h_0$& $f_2(-3A+2B+C,-B+D,-B+E)+f_4(-B+D,-B+E)+f_6(-C+E)+f_7(-D+E)$\\
 \hline
\textbf{S10} & $ \xi (t)$&$\eta _a (a)$ & $\eta_N(a,N)$& known  &  arbitrary  & arbitrary  & known  & known  &$h_0$ & $f_2(-3A+2B+C,-B+D,-B+E)+f_4(-B+D,-B+E)+f_5(-C+D,-C+E)$\\
 \hline
\textbf{S11} &$ \xi (t)$ &$\eta _a (a)$ & $\eta_N(a,N)$& known  & known  & known   & arbitrary  & arbitrary  &$h_0$ & $f_2(-3A+2B+C,-B+D,-B+E)+f_3(-B+C,-B+D,-B+E)$\\
 \hline
\textbf{S12} &$ \xi (t)$ &$\eta _a (a)$ & $\eta_N(a,N)$& known  & known  & known  & arbitrary  & arbitrary  &$h_0$ & $f(-A+B,-A+C,-A+D,-A+E)$\\
 \hline
 \hline
  \end{tabular}}
  \caption{Here we present the symmetries \textbf{S\textit{n}} of the diagram \ref{fig:M1}.  Specifically, in the last column we write down the form of the theory that is invariant under the Noether Symmetry $\X$, whose coefficients are given in the columns 2 to 10. We write \textit{arbitrary} when this coefficient is not specified, i.e. the system is solved for any form of this and \textit{known} for the cases where the form of a coefficient is known and it depends on the rest coefficients. In the first row of the table, the argument $X$ stands for $f(X) \equiv f(A,B,C,D,E)$, etc.}
  \label{tab:sym}
\end{table}

\section{Noether Symmetries of a conformally invariant theory}
\label{sec:Noether_conformal}

\begin{figure}[h!]
\begin{center}
\begin{tikzpicture}
\draw[gray, thick] (-4,0) -- (4,0);
\draw[gray, thick] (-4,0) -- (-4,-1);
\draw[gray, thick] (-6,-1) -- (-2,-1);
\draw[gray, thick] (4,0) -- (4,-1);
\draw[gray, thick] (2,-1) -- (6,-1);
\draw[gray, thick] (2,-1) -- (2,-2);
\draw[gray, thick] (1,-2) -- (3,-2);
\draw[gray, thick] (1,-2) -- (1,-3);
\draw[gray, thick] (0,-3) -- (2,-3);
\draw[gray, thick] (-2,-1) -- (-2,-2);
\draw[gray, thick] (-3,-2) -- (-1,-2);
\filldraw[black] (-4,0)  node[anchor=south] {Case 1};
\filldraw[black] (4,0)  node[anchor=south] {Case 2};
\filldraw[black] (-6,-1)  circle (2pt) node[anchor=south] {C1.a};
\filldraw[black] (-2,-1)  node[anchor=south] {C1.b};
\filldraw[black] (2,-1)  node[anchor=south] {C2.a};
\filldraw[black] (6,-1) circle (2pt) node[anchor=south] {C2.b};
\filldraw[black] (-3,-2)  circle (2pt) node[anchor=south] {C1.b.i};
\filldraw[black] (-1,-2) circle (2pt) node[anchor=south] {C1.b.ii};
\filldraw[black] (1,-2) node[anchor=south] {C2.a.i};
\filldraw[black] (3,-2) circle (2pt) node[anchor=south] {C2.a.ii};
\filldraw[black] (0,-3) circle (2pt) node[anchor=south] {C2.a.i.1};
\filldraw[black] (2,-3) circle (2pt) node[anchor=south] {C2.a.i.2};
\filldraw[black] (-6,-1)  node[anchor=north] {\textbf{S1}};
\filldraw[black] (-3,-2)  node[anchor=north] {\textbf{S2}};
\filldraw[black] (-1,-2)  node[anchor=north] {\textbf{S3}};
\filldraw[black] (0,-3)  node[anchor=north] {\textbf{S4}};
\filldraw[black] (2,-3)  node[anchor=north] {\textbf{S5}};
\filldraw[black] (3,-2)  node[anchor=north] {\textbf{S6}};
\filldraw[black] (6,-1)  node[anchor=north] {\textbf{S7}};
\end{tikzpicture}
\end{center}
\caption{As in the previous section, here we represent with ``C\textit{n.l}'' the different cases of the table \ref{tab:cases2} and with ``\textbf{S\textit{n}}'' the symmetries of the theory \eqref{point2} shown in table \ref{tab:sym2}.}
\label{diagram2}
\end{figure}
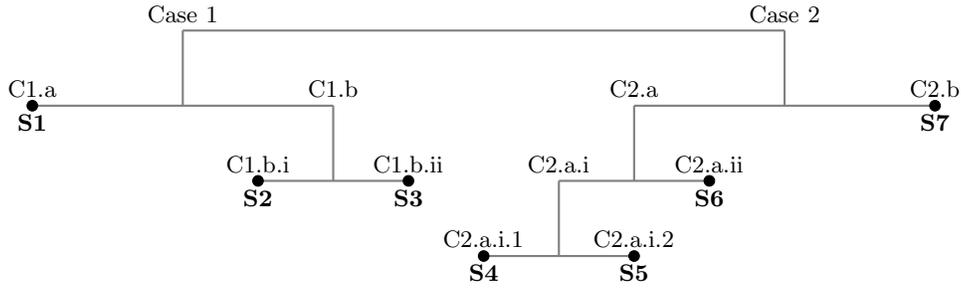

It turns out that the point-like Lagrangian \eqref{point} with the function $f$ given by \eqref{fnew}, takes the form
\begin{equation}\label{point2}
\mathcal{L} = 24 \left(b_1+3 b_3-b_4\right)\frac{ a^2 }{N^2}\dot{a}\dot{N} - 12 \left(b_1+3 \left(b_3+b_4\right)\right)\frac{ a}{N}\dot{a}^2-4 \left(3 b_1+9 b_3+b_4\right)\frac{ a^3 }{N^3}\dot{N}^2 + b_5 a^3 N \,.
\end{equation}
Apparently, there is no explicit dependence on the non-metricity scalars and the configuration space of \eqref{point2} is $\mathcal{Q} = \{ a, N\}$. Consequently, the Noether vector takes the form
\begin{equation}
\X = \xi (t,a,N) \partial _a + \eta _a(t,a,N) \partial _a + \eta _N(t,a,N) \partial _N\,.
\end{equation}
Applying the Noether condition \eqref{NoetherCondition} to the Lagrangian \eqref{point2} we obtain a system of 10 equations, which leads to 7 different cases of symmetries. These are described in the figure \ref{diagram2} and details are given in the tables \ref{tab:cases2} and \ref{tab:sym2}.

\begin{table}[!h]
\renewcommand{\arraystretch}{1.5}
\centering
\centering
\begin{tabular}{c|c}
\hline
\hline
\textbf{Cases} & \textbf{Conditions}   \\ 
\hline
\textbf{1} & $b_1 = -3 (b_3 + b_4)$ \\ 
\hline 
\textbf{2} &  $b_1 \neq -3 (b_3 + b_4)$\\
 \hline 
\textbf{C1.a} & $b_4  = 0$\\
 \hline 
 \textbf{C1.b} & $b_4 \neq 0$\\
 \hline 
 \textbf{C1.b.i} & $b_5 = 0$\\
 \hline 
 \textbf{C1.b.ii} & $b_5 \neq 0$\\
 \hline 
 \textbf{C2.a} & $b_5 \neq 0 $\\
   \hline 
 \textbf{C2.b} & $b_5 = 0$\\
   \hline 
 \textbf{C2.a.i} & $b_1  \neq -3 b_3$\\
   \hline 
 \textbf{C2.a.ii} & $b_1 = -3 b_3$\\
   \hline 
 \textbf{C2.a.i.1} & $b_4 = 0$ and $\eta_{N1} \neq 0$\\
   \hline
 \textbf{C2.a.i.2} & $b_4 \neq 0$ and $\eta_{N1} = 0$\\
 \hline
 \hline
  \end{tabular}
  \caption{Here we present the different cases shown on the figure \ref{diagram2}.}
  \label{tab:cases2}
\end{table}


\begin{table}[!h]
\renewcommand{\arraystretch}{1.5}
\centering
\centering
\resizebox{20cm}{!}{
\begin{tabular}{c|c|c|c|c|c}
\hline
\hline
\textbf{Symmetry} & $\xi (t,a,N)$ & $\eta _a(t,a,N)$ & $\eta _N(t,a,N)$ & $h(t,a,N)$ & $f(A,B,C,D,E) $  \\ 
\hline
\hline
\textbf{S1} & $ \xi (t)$ & $-\frac{a \eta _Ν(t,a,Ν)}{3 N}+\frac{h(t)}{3 a^2 b_5 N}-\frac{1}{3} a \xi'(t)$& $\eta _N (t,a,N)$& $h(t) $& $-3 b_3 A + b_2 B + b_3 C -\left(b_2-4 b_3\right) D - 2 b_3 E+b_5$\\
\hline
\textbf{S2} & $ c_1 t - c_2 t^2 +c_3$ & $\frac{-c_7  N t}{96 a^2 b_4}+\frac{a^3 \left((-c_2 +c_5) t+c_1+c_4-c_6\right)+3 c_9 N}{3 a^2}+\frac{1}{3} a \left(c_6-c_2 t\right) \ln (N)$& $N \left(\left(c_6-c_2 t\right) \ln (N)+c_5 t+c_4\right)$& $-\frac{32 b_4 a^3  \left(c_5-c_2 \ln(N)\right)}{N}+c_8 \ln(N)+c_7 $& $-3 (b_3 + b_4) A + b_2 B+ (b_3+b_4)C - (b_2-4 b_3+12 b_4)D - (2 b_3-6 b_4)E$\\
\hline
\textbf{S3} & $ \xi_0$ & 0 &  0 & $h_0/2 $& $-3 A (b_3+b_4)+b_2 B+\left(b_3+b_4\right) C-\left(b_2-4 b_3+12 b_4\right) D - \left(2 b_3-6 b_4\right)E +b_5$\\
\hline
\textbf{S4} & $c_1 t+c_2$ & $-\frac{8 \sqrt{2} c_3 a^{1/4}  \left(\left(b_1+3 b_3\right) N\right){}^{7/4}}{3 N}$& $\frac{8 \sqrt{2} c_3 a^{1/4}  \left(\left(b_1+3 b_3\right) N\right){}^{7/4}}{3 N}- c_1 N$ &  $h_0 $& $b_1 A+b_2 B+b_3 C+\left(4 b_1-b_2+16 b_3\right) D-2  \left(b_1+4 b_3\right)E +b_5$\\
\hline
\textbf{S5} & $ c_1 t+c_2$ & 0 & $-c_1 N$& $h_0 $& $b_1 A+b_2 B+(b_3+b_4) C+\left(4 b_1-b_2+16 b_3\right) D-2  \left(b_1+4 b_3\right)E +b_5$\\
\hline
\textbf{S6} & $ c_1 t+c_2$ & 0 & $-c_1 N$& $h_0 $ & $-3 b_3 A+b_2 B+\left(b_3+b_4\right) C-\left(b_2-4 b_3\right) D-2 b_3 E +b_5$\\
\hline
\textbf{S7} & $ \xi (t)$ & $\frac{1}{6} a \left(-6 a^3 N \eta _a'\left(a^3 N\right)+3 \eta _a\left(a^3 N\right)+\xi '(t)\right)$ & $\frac{1}{2} a \left(6 a^3 N \eta _a'\left(a^3 N\right)+3 \eta _a\left(a^3 N\right)-\xi '(t)\right)$ & $h_0$ & $-3 b_3 A+b_2 B+\left(b_3+b_4\right) C-\left(b_2-4 b_3\right) D-2 b_3 E$\\
 \hline
 \hline
  \end{tabular}}
  \caption{This table shows the symmetries of the theory \eqref{point2} that are presented in figure \ref{diagram2}. Specifically, in the last column we present the exact form of the theory and in the rest of the columns the coefficients of the Noether vector, as well as the value of the function $h(t,a,N)$ in the right hand sind of \eqref{NoetherCondition}. $b_i$'s and $c_i$'s are constants.}
  \label{tab:sym2}
\end{table} 

The conserved quantities of each case will be given by
\begin{equation}
I = \xi \left( \dot{a}\frac{\partial L}{\partial \dot{a}} +\dot{N}\frac{\partial L}{\partial \dot{N}} - L \right) - \eta _a \frac{\partial L}{\partial a}- \eta _N \frac{\partial L}{\partial N} + h\,.
\end{equation}

Regarding the cosmological solutions now, we can consider e.g.  the model 1.b.i, i.e. the symmetry \textbf{S2}, for which the form of the function in the action is
 \begin{equation}
 f(A,B,C,D,E) = -3 (b_3 + b_4) A + b_2 B+ (b_3+b_4)C - (b_2-4 b_3+12 b_4)D - (2 b_3-6 b_4)E \,,
 \end{equation}
and the point-like Lagrangian \eqref{point2} becomes
\begin{equation}
L = \frac{32 b_4 a^2}{N^3} \left(a \dot{N}^2-3 N \dot{a} \dot{N}\right)\,.
\end{equation}
The Euler-Lagrange equations for the above Lagrangian read
\begin{gather}
\frac{96 b_4 a^2}{N^3}\left(N \ddot{N}-\dot{N}^2\right) = 0\,,\\
-\frac{32 b_4 a}{N^4} \left[2 a N \left(3 \dot{a} \dot{N}+a \ddot{N}\right)-3 N^2 \left(a \ddot{a}+2 \dot{a}^2\right)-3 a^2 \dot{N}^2\right] = 0 \,.
\end{gather}
The above system can be easily solved to give
\begin{equation}
N(t) = N_1 e ^{N_2 t}\quad \text{and}\quad a(t) = a_1 \left(3(t-a_2)\right)^{1/3} e^{\frac{N_2}{3}\left(t-3a_2\right)}\,, 
\end{equation}
where $a_1,\,a_2,\,N_1$ and $N_2$ are constants of integration. We see that, for specific choices of the integration constants, we can have both power-law and de-Sitter-like solutions. The same approach to find out exact cosmological solutions can be worked out for the other cases above.

\section{Discussion and Conclusions}
\label{sec:Discussion}

Modifications of gravity are widely studied   in different contexts. However, modifications of the alternative formulations of GR, i.e. TEGR and STEGR,  lead to different phenomenology due to the difference in the underlying geometry. In this paper, we studied a general theory containing all five even-parity, quadratic scalars of non-metricity, $f(A,B,C,D,E)$. Specifically, in order to select specific models of such a general theory, apart from observational constraints, we can apply some theoretical ones as well. Here we considered the Noether Symmetry Approach in  cosmological minisuperspaces generated from  $f(A,B,C,D,E)$ and we classified all those models that are invariant under point-transformations. 
In addition to that, we showed that these symmetries can be used in order to calculate the zero-order invariants of the theory, reduce its dynamics and find exact solutions in the spacetime under consideration. 

For completeness, we studied a scale- and conformally- invariant theory, i.e. \eqref{fnew} and \eqref{point2}, and we found what are the cases that show Noether symmetries. According to this result,  these models can be exactly  solved  giving rise to physically relevant cosmological solutions. In general, they can be useful for a more extensive cosmographic analysis matching viable cosmological models with observational data \cite{Capozziello:2019cav, Dunsby:2015ers}.

The forthcoming perpective is to check if those models that are invariant under point-transformations in the cosmological minisuperspace, present a similar behavior in other spacetimes as well, e.g. with spherical or cylindrical symmetry. In such a case, the theoretical hints on the viability of the models would be  more powerful. In general,  if symmetries exist, one is able to find out exact black hole solutions for those theories and to study the physical relevance (see for example \cite{Bahamonde:2016jqq, Paliathanasis:2014iva, Capozziello:2012iea, Capozziello:2007wc}).

\begin{acknowledgments}
This article was based upon work from CANTATA COST (European Cooperation in Science and Technology) action CA15117, EU Framework Programme Horizon 2020. KFD, whose work was partially done at the NORDITA institute in Stockholm and at the University of Salamanca in Spain, is grateful to TSK and to Jos\'e Beltr\'a{}n Jim\'e{}nez for the hospitality. SC acknowledges INFN ({\it iniziativa} specifica QGSKY) for partial support. TSK  acknowledges support from the Estonian Research Council through the Personal Research Funding project PRG356 ``Gauge Gravity'' and from the European Regional Development Fund through the Center of Excellence TK133 ``The Dark Side of the Universe''.
\end{acknowledgments}


\end{document}